\documentclass[pra,aps,twocolumn,showpacs,amsmath,amssymb,floatfix,10pt]{revtex4-1}
\input{epsf}

\usepackage{graphicx}
\usepackage{dcolumn}
\usepackage{bm}

\begin{document}
\title{Decoherence of interacting Majorana modes}
\author{H. T. Ng}
\thanks{Corresponding author}
\email{hotsangng@tsinghua.edu.cn}
\affiliation{Center for Quantum Information, Institute for Interdisciplinary Information Sciences, Tsinghua University, Beijing 100084, P. R. China}

\date{\today}

\begin{abstract}
We study the decoherence of Majorana modes of a 
fermion chain, where the fermions interact
with their nearest neighbours. We investigate the effect
of dissipation and dephasing on the Majorana modes
of a fermionic chain. The dissipative and dephasing
noises induce the non-parity- and parity-preserving transitions between the eigenstates of the system, respectively. Therefore, these two types of noises
lead to the different decoherence mechanisms. 
In each type of noise, we discuss
the low- and high-frequency regimes to describe the 
different environments. We numerically calculate
the dissipation and dephasing rates in the presence of long-range interactions. We find that the decoherence rate of interacting Majorana modes is different to that of non-interacting modes. We show the examples that the long-range interactions can reduce the decoherence rate. It is advantageous to the potential applications of quantum information processing.
\end{abstract}


\maketitle

\section{Introduction}
Majorana fermions are exotic particles \cite{Wilczek} which show non-abelian statistics \cite{Read,Ivanov,Alicea}. Indeed, non-abelian statistics is necessary for performing topological quantum computation \cite{Nayak}
which is a kind of fault-tolerant quantum computation.
Thus, the study of Majorana fermions is of fundamental 
importance and also it is useful to the applications of quantum information processing (QIP). 

Kitaev predicted that an unbound pair of Majorana fermions \cite{Kitaev} exhibits at the two ends of a spin-polarized one-dimensional (1D) superconductor.
This provides a promising way to realize Majorana fermions.
Recently, a number of methods has been proposed to simulate Majorana fermions in a 1D system such as by using a semiconductor nanowire \cite{Lutchyn,Oreg} and cold atoms in an optical lattice \cite{Jiang,Kraus}.

Decoherence severely hinders the performance of QIP applications which rely on quantum coherence \cite{Nielsen}. 
The various approaches have been proposed to 
combat against decoherence such as quantum error correction \cite{Shor,Steane} and dynamical decoupling techniques \cite{Viola,Facchi}, etc. 
Remarkably, Majorana fermions are robust against local perturbations \cite{Kitaev2} due to a large energy gap from the two degenerate ground states. It is believed that they can be exploited without further protection.  Still, they suffer from decoherence.
Recently, decoherence of Majorana modes has been studied in more detail \cite{Goldstein,Budich,Schmidt,Cheng,Rainis,Ho}. The noises sources from the different physical settings have also been discussed \cite{Schmidt,Cheng,Rainis}. 

In addition, the effects of long-range interactions between fermions
on the Majorana modes \cite{Stoudenmire,Gangadharaiah,Sela,Hassler,Thomale,Ng} have recently been studied. 
The long-range interactions can broaden the range of parameters for exhibiting Majorana fermions \cite{Stoudenmire,Ng}. It is natural to ask the effect of long ranged interactions on decoherence of the Majorana modes. In this paper, we study the decoherence rate of Majorana modes of a chain of spinless fermions in the presence of long-range interactions between fermions.
Our study is helpful to understand the relationship between interactions and the decoherence
properties in a many-body system.

We study the two typical noises in the system, where they are dissipation and dephasing, respectively. These
two types of noises are widely studied in the context 
of open quantum problems and also they are two main forms
of decoherence occurring in quantum computing \cite{Ladd}.
Dissipation and dephasing lead to the different decoherence mechanisms of Majorana modes. Dissipation induces the non-parity
preserving transitions between the eigenstates of the system while dephasing gives rise to parity preserving transitions.

Moreover, we investigate the low- and high-frequency
noises to describe the different types of environment.
The frequency domain of the low-frequency noise spectrum is much lower than the transition frequency of the two degenerate ground states and their first excited states.
For example, the low-frequency noise can be described by the $1/f$-noise \cite{Paladino} which commonly occurs in the solid-state devices.
On the other hand, the high-frequency noise is to describe the environment in which the frequency domain
of the noise spectrum is comparable to the transition
frequencies between the different eigenstates. 
We consider the high-frequency baths to be Markovian in this paper.

We show the examples that the long ranged interactions 
between fermions can reduce the decoherence rates. 
In fact, the dissipation and dephasing rates depend on the collective properties of fermions which can be changed by the interactions between the fermions. As a result, \textit{long ranged interactions can change the decoherence properties of Majorana modes.}
In this way, the coherence time of the Majorana modes can be prolonged by appropriately choosing the interaction parameters. It may be useful for Majorana-based applications \cite{Alicea,Nayak,Mazza}.

\section{System}
Majorana modes occur in a spin-polarized 1D superconductor \cite{Kitaev}. This 1D superconductor can be described by a chain of spinless fermions with an open boundary condition. The Hamiltonian of this fermionic system is given by, $(\hbar=1)$,
\begin{eqnarray}
H_{\rm 1D}&=&\sum_j\big[-w(c^\dag_jc_{j+1}+c^\dag_{j+1}c_j)
+\Delta(c_jc_{j+1}+c^\dag_{j+1}c^\dag_j)\big]\nonumber\\
&&-\mu\sum_j\Big(c^\dag_jc_j-\frac{1}{2}\Big),
\end{eqnarray}
where $c_j$ and $c^\dag_j$ are annihilation 
and creation fermionic operators at site $j$.
The parameters $w$, $\Delta$ and $\mu$ are the 
tunneling strength, superconducting gap and chemical potential, respectively.

We consider the fermions to be interacted with
their nearest neighbors. The Hamiltonian, describes
long-range interaction \cite{Gangadharaiah}, is written as,
\begin{eqnarray}
H_{\rm U}&=&U\sum_j\Big(c^\dag_jc_j-\frac{1}{2}\Big)\Big(c^\dag_{j+1}c_{j+1}-\frac{1}{2}\Big),
\end{eqnarray}
where $U$ is the repulsive interaction strength between
the nearest neighbours. 

A fermionic chain can be mapped onto a spin chain by applying the Jordan-Wigner transformation \cite{Kitaev2}.
The fermionic operators are related to spin-half operators via the Jordan-Wigner transformation as follows:
\begin{eqnarray}
c_j&=&(-1)^{j-1}\prod^{j-1}_{k=1}\sigma^z_k\sigma^-_j,\\
c^\dag_j&=&(-1)^{j-1}\prod^{j-1}_{k=1}\sigma^z_k\sigma^+_j,\\
\label{occupation_number}
c^\dag_jc_j&=&\frac{1}{2}(\sigma^z_j+1),
\end{eqnarray}
where $\sigma^{\pm}_j$ and $\sigma^z_j$ are the Pauli spin operators at site $j$.
The Hamiltonian $H=H_{\rm 1D}+H_{\rm U}$ of the system can be recast as 
\begin{eqnarray}
\label{spinHam}
H&=&\sum_j[w(\sigma^+_j\sigma^-_{j+1}+\sigma^-_{j}\sigma^+_{j+1})
+\Delta(\sigma^+_{j}\sigma^+_{j+1}+\sigma^-_j\sigma^-_{j+1})]\nonumber\\
&&-\frac{\mu}{2}\sum_j\sigma^z_j+\frac{U}{4}\sum_j\sigma^z_j\sigma^z_{j+1}.
\end{eqnarray}
The quantum simulation of the Ising spin chain with
the transverse field by using trapped ions has recently been proposed \cite{Mezzacapo}.  

This 1D system possesses the $\mathbb{Z}_2$ symmetry.
The parity operator $P$ can be defined as
$(-1)^{\sum^N_j{a^\dag_j{a_j}}}$ and $\prod^N_{j=1}\sigma^z_j$ for a fermionic chain and a spin chain, respectively.
Therefore, each eigenstate has a definite parity. It
is either to be $P=1$ (even) or $P=-1$ (odd). 

\section*{Majorana fermions}
Majorana operators can be defined as \cite{Kitaev,Kitaev2}
\begin{eqnarray}
c_{2j-1}&=&a^\dag_j+a_j,~~c_{2j}=i(a^\dag-a_j).
\end{eqnarray}
The Majorana operators satisfy the anti-commutation rules, and also they are Hermitian operators.
In fact, the Hamiltonian of a fermonic chain can be expressed in terms of Majorana operators \cite{Kitaev,Kitaev2}.
A pair of unbound Majorana fermions exhibit at the ends of a chain and the remaining Majorana fermions are
bounded in pair \cite{Kitaev,Kitaev2}. The pair of unbound Majorana fermions
(Majorana modes) are shown when the system has the two-fold ground-state degeneracy, where the two degenerate ground states have the different parities. 

The Majorana modes can exhibit even
if the fermions interact with their nearest neighbours \cite{Stoudenmire,Ng}. This can be indicated by examining the ground-state degeneracy. We calculate the energy
difference between the two ground states with the 
different parities. It can be defined as \cite{Stoudenmire}
\begin{eqnarray}
\label{energydiff}
\Delta{E}&=&|E^e_1-E^o_1|,
\end{eqnarray}
where $E^e_1$ and $E^o_1$ are the ground-state
eigen-energies in the even- and odd-parities,
respectively. If $\Delta{E}$ is zero, then
the system supports the Majorana modes \cite{Stoudenmire}.

\begin{figure}[ht]
\centering
\includegraphics[height=8.0cm]{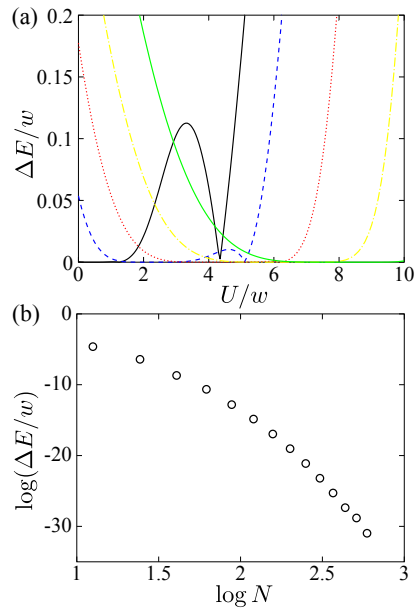}
\caption{ \label{Energy_gap1} (Color online) 
In (a), energy gap $\Delta{E}$ versus interaction strength $U$,
for $N=12$ and $\mu=w$. The different interaction strengths
$\Delta$ are denoted by the different lines: $\Delta=w$ (black solid), $2w$ (blue dashed), $3w$ (red dotted), $4w$ (yellow dash-dotted) and $5w$ (green solid), respectively.
In (b), log-log plot of energy gap $\Delta{E}$ versus $N$, for $\mu=w$, $\Delta=5w$ and $U=8w$.
}
\end{figure}

We numerically 
solve the Hamiltonian in Eq.~(\ref{spinHam})
by using exact diagonalization. In Fig.~\ref{Energy_gap1}(a), we plot the energy difference $\Delta{E}$ as a function of interaction strength $U$, for the different
interaction strengths $\Delta$. The zero energy gap is
shown, this implies that the Majorana modes exist.
When $\Delta$ increases, the broader range of interaction
strength $U$ can be obtained. We also study the relation of the energy gap and the size $N$ of system. In Fig.~\ref{Energy_gap1}(b), we plot $\Delta{E}$ verus $N$ in the logarithmic scale. The energy gap exponentially decreases
as the size $N$. This shows that the feature of topological degeneracy \cite{Kitaev2}.

\section{Phase diagram}
To understand the ground-state properties of the system, we briefly discuss the phase diagram. To facilitate our discussion, we recast the 
Hamiltonian in Eq.~(\ref{spinHam}) as
\begin{eqnarray}
\label{XYZ}
H_{\rm XYZ}&=&\frac{w}{2}\sum_j\bigg[\Big(1+\frac{\Delta}{w}\Big)\sigma^x_j\sigma^x_{j+1}+\Big(1-\frac{\Delta}{w}\Big)\sigma^y_{j}\sigma^y_{j+1}\nonumber\\
&&+\frac{U}{2w}\sigma^z_j\sigma^z_{j+1}\bigg]-\frac{\mu}{2}\sum_j\sigma^z_j.
\end{eqnarray}
Indeed, it is the $XYZ$ model \cite{Sela,Pinheiro}.
Note that the system is invariant if the sign of $\mu$
is changed, i.e., $\mu\rightarrow{-\mu}$. This can be
seen by transforming the spin operators $\sigma^{x,z}_j$
into $-\sigma^{x,z}_j$. The Hamiltonian $H_{\rm XYZ}$ in Eq.~(\ref{XYZ}) remains unchanged.

\begin{figure}[ht]
\centering
\includegraphics[height=4.50cm]{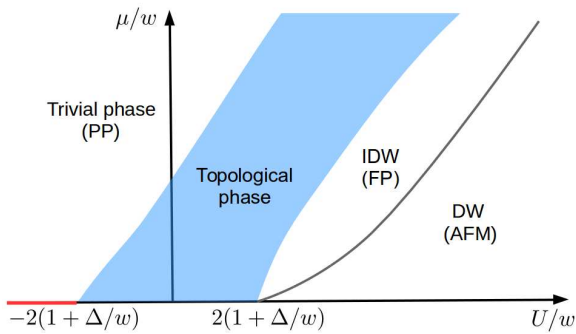}
\caption{ \label{SD_XYZ} (Color online) 
Schematic of phase diagram of the $XYZ$ model (see, e.g., \cite{Sela,Hassler,Thomale,Pinheiro}).
The red line is marked for the transition when $\mu=0$
and $U<-2(1+|\Delta|/w)w$.
}
\end{figure}

\begin{figure}[ht]
\centering
\includegraphics[height=10.0cm]{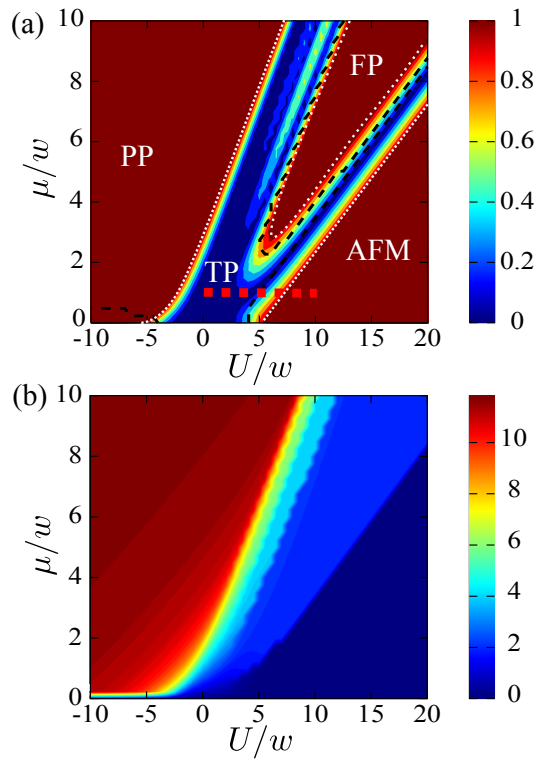}
\caption{ \label{Exact_XYZ1} (Color online) 
Contour plots of $\Delta{E}$ and $M$ versus interaction strengths $\mu$ and $U$ in (a) and (b), respectively, for
$N=12$ and $\Delta=w$.
In (a), the black dashed lines are marked to indicate
that the two ground states occur in the same parity.
The red horizontal dotted line is marked for the parameters we discussed in the subsequent figures.
The different phases are labelled and the white 
dotted lines are used for showing the phase region.
}
\end{figure}

The phase diagram of the $XYZ$ model is known \cite{Sela,Hassler,Thomale,Pinheiro}. Let us
briefly discuss their results.
The schematic of phase diagram as a function of $\mu$ and $U$ is shown in Fig~\ref{SD_XYZ}. This system has the four different phases. They are trivial, topological, density-wave (DW) and incommensurate density-wave (IDW) phases.
The topological phase can be found by examining
the energy difference $\Delta{E}$ in Eq.~(\ref{energydiff}) between the two ground states with the different parities \cite{Stoudenmire,Thomale}. The DW and IDW phases can be found when the two ground states occur in the same parity \cite{Thomale}. The DW phase is also called the anti-ferromagnetic (AFM) in which the total magnetization becomes zero in the $z$ direction \cite{Pinheiro}. But the IDW phase, which is termed as floating phase \cite{Sela,Hassler,Pinheiro}, has a finite magnetization. Also, at the zero magnetic field ($\mu=0$), the system is characterized by a ferromagnetic (FM) phase \cite{Pinheiro} for large negative $U$. When the magnetic field becomes large, the system is in a trivial (PP) phase with a large magnetization which depends on the direction of the magnetic field. There is a transition \cite{Pinheiro} between them when $U$ is less than $-2(1+|\Delta|/w)w$.

We examine the ``finite-size'' phase diagram by studying
$\Delta{E}$ and the total magnetization $M=\sum_j\langle\sigma^z_j\rangle$ in the $z$ direction. In Fig.~\ref{Exact_XYZ1}(a), the contour plot of $\Delta{E}$ is plotted as a function of $\mu$ and $U$. The topological phase (TP) can be indicated when $\Delta{E}=0$, i.e. the deep blue region in Fig.~\ref{Exact_XYZ1}(a). Indeed, the
topological phase can be described by the two N\'{e}el states
in the $x$-direction. A more detailed discussion can be found in supplementary information.
When the two ground states occur in the same parity, the DW and IDW phases can be distinguished from the topological phase in Fig.~\ref{Exact_XYZ1}(a). Also, the transition between the FM and PP phases at zero $\mu$ can also be indicated in Fig.~\ref{Exact_XYZ1}(a). In addition, we plot the total
magnetization $M$ versus $\mu$ and $U$ in Fig.~\ref{Exact_XYZ1}(b). The trivial (PP) and DW (AFM) phases can be clearly shown. But the transition between the topological phase and IDW phase cannot
be distinguished by this method \cite{Pinheiro}. 
By comparing the energy gap and its parity and also the magnetization, we are able to determine the phase which
is labelled in Fig.~\ref{Exact_XYZ1}(a). The transitions
between the different phases cannot be manifestly shown due to the relatively small size of the system.

\begin{figure}[ht]
\centering
\includegraphics[height=6.00cm]{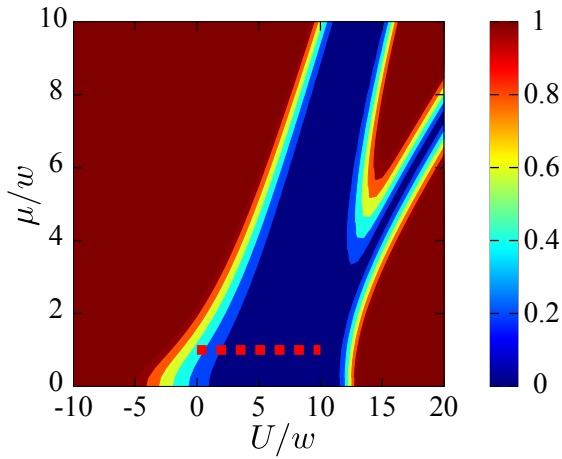}
\caption{ \label{Exact_XYZ2} (Color online) 
Contour plots of $\Delta{E}$ versus interaction strengths $\mu$ and $U$, for $N=12$ and $\Delta=5w$. The red dotted horizontal line is marked for the parameters used
in the subsequent figures.
}
\end{figure}

In Fig.~\ref{Exact_XYZ2}, we show the contour plot of $\Delta{E}$ versus $\mu$ and $U$ with a larger $\Delta=5w$. In this case, the region of nearly zero $\Delta{E}$ becomes larger than that in Fig.~\ref{Exact_XYZ1}(a) since $U$ increases. This means that the topological phase can be obtained with a wider range of parameters. However, the topological phase tends to shift to the right-hand side and it is smaller than that of the schematic phase diagram in Fig.~\ref{SD_XYZ} due to the finite-size effect.

\section{Decoherence}
We consider the fermions to be coupled to an environment.
This causes decoherence of the Majorana modes. We study the two different types of 
noises which are dissipation and dephasing,
respectively.

In general, the total Hamiltonian, which includes the system and bath and their interactions, can be written as
\begin{eqnarray}
H_t&=&H+H_B+H_{BI},
\end{eqnarray}
where $H$, $H_B$ and $H_{BI}$ are
the Hamiltonians of the system, bath and 
system-bath interactions, respectively.
It is convenient to express
the Hamiltonian $H_t$ in terms of the system's eigenstates, i.e.,
\begin{eqnarray}
\label{EsysHam}
H&=&\sum_nE^\alpha_n|n\rangle_\alpha{}_\alpha\langle{n}|,
\end{eqnarray}
where $E^\alpha_n$ is the eigen-energy of the $n$-th
eigenstate $|n\rangle_\alpha$ of the system in the even $(\alpha=e)$ and odd $(\alpha=o)$ parities.
In the interaction picture, the Hamiltonian
$H_{BI}$ can be written in terms of the eigenstate $|n\rangle_\alpha$ as
\begin{eqnarray}
H_{BI}(t)\!&=&\!\!\sum_{\alpha,\beta,n,m}|{n}\rangle_\alpha{}_\alpha\langle{n}|H_{BI}|{m}\rangle_\beta{}_\beta\langle{m}|,\\
&=&\!\!\sum_{\alpha,\beta,n,m,j}\!g_j{}_\alpha\langle{n}|s_j|{m}{\rangle}_{\beta}e^{i(E^\alpha_n-E^\beta_m)t}B_j(t)|n\rangle_\alpha{}_\beta\langle{m}|,\nonumber\\
\end{eqnarray}
where $g_j$ is the system-bath coupling strength, $s_j$ and $B_j(t)$ are the system and bath operators at site $j$, and $\alpha,\beta=e$ and $o$. Here we study the eigenstates of a spin chain which can be easier to numerically implement.

For the low-frequency noise, 
we consider the frequency domain of the noise spectrum to be much lower than
the transition frequency between the degenerate ground states and their first excited states. However,
the two degenerate ground states are still subject
to low-frequency noise. 

\begin{figure}[ht]
\centering
\includegraphics[height=4.2cm]{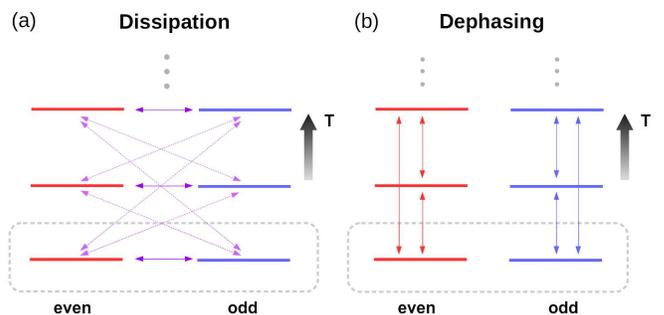}
\caption{ \label{Decoherence_induced_transition} (Color online)
Transitions between eigenstates via dissipation in (a)
and dephasing in (b). In (a), dissipation induces the transitions between the eigenstates with the different parities. In (b), dephasing induces the transitions in the same parity. In both cases, transitions between the two degenerate states occur via low-frequency noise, and transitions between higher excited states occur through high-frequency noise at finite temperature.}
\end{figure}

In the case of high-frequency
noise, the frequency domain of the noise spectrum
is comparable to the transition frequencies between the
different eigenstates. We assume that the coupling
between the system and bath is weak so that the
Born-Markovian approximation can be applied.
At zero temperature, the system maintains in the two degenerate ground states. We have also assumed that 
the coupling between the two degenerate ground states and the bath is zero for this environment.
However, the bath will induce the transitions between the degenerate ground states and higher excited states at finite temperature. In the subsequent discussion, we will study the low- and high-frequency regimes in the different types of noises.

\subsection{Dissipation}
In this subsection, we discuss the effect of dissipation on the Majorana modes. The Hamiltonian of system-bath interaction, which describes the dissipation, is of the form:
\begin{eqnarray}
H_{BI}&=&\sum_jg_j(c^\dag_j+{c_j})B_j,
\end{eqnarray}
where $g_j$ and $B_j$ are the system-bath coupling
strength and the bath operator, respectively.
Here each fermion independently couples to a
fermionic bath.
Such dissipation noise leads to transitions between the eigenstates in the different parities.
Transitions between the eigenstates in the different parities
is shown in Fig.~\ref{Decoherence_induced_transition}(a).

\begin{figure}[ht]
\centering
\includegraphics[height=6.0cm]{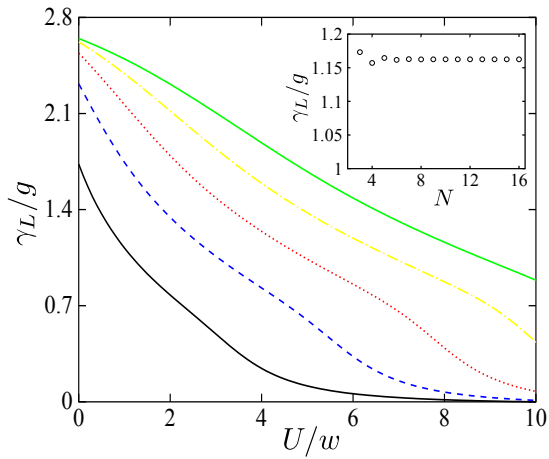}
\caption{ \label{Low_freq_dissipation} (Color online) 
Parameter $\gamma_L$ versus interaction strength $U$,
for $N=12$ and $\mu=w$. The different interaction strengths
$\Delta$ are denoted by the different lines: $\Delta=w$ (black solid), $2w$ (blue dashed), $3w$ (red dotted), 
$4w$ (yellow dash-dotted) and $5w$ (green solid), respectively. In the inset, the parameter $\gamma_L$ versus $N$, for $\mu=w$, $\Delta=5w$ and $U=8w$. 
}
\end{figure}

\subsubsection{Low-frequency noise}
Here we consider the low-frequency noise to be
dominant.
The frequency domain of the noise spectrum is much 
lower than the transition frequency between 
the two degenerate ground states 
and their first excited states.
The Hamiltonian, describes the interaction
between the two degenerate ground states and the bath, can be written as
\begin{eqnarray}
H_{BI}&=&\sum_jg_jC^{11}_j|1\rangle_e{}_o\langle{1}|e^{i\Delta^{eo}_{11}t}B(t)+H.c.,
\end{eqnarray}
where $C^{11}_j={}_e\langle{1|c_j+c^\dag_j}|1\rangle_o$, $\Delta^{eo}_{11}=E^e_1-E^o_1$,
and $B(t)$ is a time-dependent bath operator. Here
$\Delta^{eo}_{11}$ is very close to zero. It should 
be noted that the the dissipation does not
cause the energy damping to the two ground states in the
low-frequency noise, but it leads to decoherence.

We assume that the system-bath coupling strengths $g_l\approx{g}$ are nearly equal. The coupling strength between the two ground states and the bath is given by
\begin{eqnarray}
\gamma_L=g\Big|\sum_jC^{11}_j\Big|.
\end{eqnarray}
The decoherence rate is closely related to the parameter $\gamma_L$. In fact, the decoherence rate also depends
on the explicit property of the noise spectrum \cite{Paladino}. For example, we consider $1/f$ noise
which can be described by the spin fluctuator model.
The decoherence rate is proportional to the ratio
of $\gamma_L$ to $\gamma_f$, where $\gamma_f$ is 
the switching rate of spin fluctuator. Therefore,
the parameter $\gamma_L$ plays an important role to
describe the decoherence effect.
Here we investigate the parameter $\gamma_L$ only.
This parameter $\gamma_L$ can reflect how 
strong the decoherence effect is.
In Fig.~\ref{Low_freq_dissipation}, we plot
$\gamma_L$ versus 
the interaction strength $U$, for the different
strengths $\Delta$. The parameter $\gamma_L$ decreases
as $U$ increases. This means that the interactions
between fermions can reduce the decoherence rate 
in the low-frequency regime.
In addition, we plot $\gamma_L$ versus $N$ in the inset of Fig.~\ref{Low_freq_dissipation}. The parameter
$\gamma_L$ is nearly constant when the system $N$
grows. We briefly discuss why this parameter $\gamma_L$ does not depend on $N$ in supplementary information.

\subsubsection{High-frequency noise}
Now we study the effect of 
dissipation on the Majorana modes, where the 
frequency domain of the noise spectrum is comparable to
the transition frequencies between the
different eigenstates. We assume that this high-frequency noise does not affect the dynamics between
the two degenerate ground states, where their 
transition frequency is nearly zero.
We consider that 
the environment can be modelled by a bath of fermions. In the interaction picture,
the Hamiltonian of system-bath coupling 
can be written as
\begin{eqnarray}
H_{BI}&=&\sum_{j,k}\sum_{n,m}g_{j_k}C^{nm}_j|n\rangle_{e}{}_{o}\langle{m}|e^{i\Delta^{eo}_{nm}t}(b_{j_k}e^{-i\omega_{j_k}t}\nonumber\\
&&+b^\dag_{j_k}e^{i\omega_{j_k}t})+{H.c.},
\end{eqnarray}
where $C^{nm}_j={}_e\langle{n}|c_j+c^\dag_j|{m}\rangle_o$ and $\Delta^{eo}_{nm}=E^e_n-E^o_m$.
The coupling strength $g_{j_k}$ is much smaller than
$|\Delta^{eo}_{nm}+\omega_{j_k}|$, where $\Delta^{eo}_{nm}\geq{0}$ and $n>m$. Therefore, we can apply the rotating-wave-approximation (RWA) to ignore the fast-oscillating terms.
The Hamiltonian can be written as
\begin{equation}
H_{BI}=\sum_{j,k}\sum_{nm}g_{j_k}C^{nm}_j|n\rangle_e{}_o\langle{m}|b_{j_k}e^{i(\Delta^{eo}_{nm}-\omega_{j_k})t}+{H.c.}.
\end{equation}

We assume that the Born-Markovian approximation
can be applied to this system. The master equation
can be derived \cite{Beaudoin} in the dressed-state picture which can provide the correct steady state
even for a strongly interacting system.
The master equation, which describes the dissipation, can be written as \cite{Beaudoin}
\begin{eqnarray}
\label{mastereqtn}
\dot{\rho}&=&-i[H,\rho]+\sum_{nm}\Gamma_{nm}[1-\bar{n}_f(\Delta^{eo}_{nm})]\mathcal{L}(|m\rangle_{e}{}_{o}\langle{n}|)\rho\nonumber\\
&&+\sum_{nm}\Gamma_{nm}\bar{n}_f(\Delta^{eo}_{nm})\mathcal{L}(|n\rangle_{e}{}_o\langle{m}|)\rho,
\end{eqnarray}
where $\Gamma_{nm}=2{\pi}d(\Delta^{eo}_{nm}){g^2}\sum_j|C^{nm}_j|^2$ and
$d(\Delta^{eo}_{nm})$ is the density of states, $g_{j_k}\approx{g}$
and $n>m$. The parameter $\bar{n}_f(\Delta^{eo}_{nm})=[\exp(\hbar\Delta^{eo}_{nm}/k_BT)+1]^{-1}$ is the mean occupation number for fermions at the frequency $\Delta^{eo}_{nm}$,
where $k_B$ is the Boltzmann constant and $T$ is the 
temperature.
The superoperator $\mathcal{L}(\rho)$ is of the Lindblad form as \cite{Breuer}
\begin{eqnarray}
\mathcal{L}(\rho)&=&s{\rho}s^\dag-\frac{1}{2}(\rho{s^\dag{s}}+{s^\dag{s}}\rho),
\end{eqnarray}
where $s=|m\rangle\langle{n}|$ and $m<n$.

The master equation in Eq.~(\ref{mastereqtn}) 
is valid if there is no degeneracy between the transitions \cite{Beaudoin}.
We assume that there is no degeneracy
between the transitions in deriving the master equation in Eq.~(\ref{mastereqtn}). The energy difference $\Delta^{eo}_{nm}$ is large enough and the
system-bath coupling $g_{j_k}$ is sufficiently weak.
Therefore, the RWA can be applied to the master equation
to ignore the fast-oscillating terms \cite{Beaudoin}.
Although it may encounter the accidental degeneracy
of the transitions between the higher excited states,
we can ignore those transitions within the coherence
time of the degenerate ground states at low temperature. The master
equation can give a reasonably good approximation
to describe the dynamics of the Majorana modes.

\begin{figure}[ht]
\centering
\includegraphics[height=6.5cm]{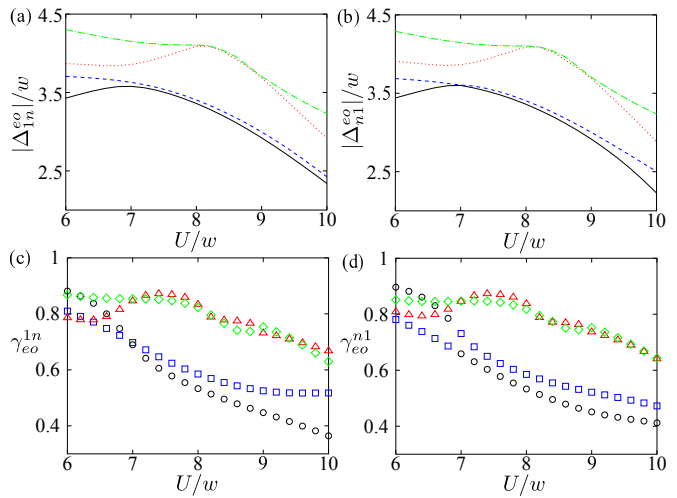}
\caption{ \label{fig_high_freq_dissipation} (Color online) 
Energy differences versus $U$ in (a)
and (b). The energy differences $|\Delta^{eo}_{1n}|=|E^{e}_{1}-E^{o}_n|$ and $|\Delta^{eo}_{n1}|=|E^{o}_{1}-E^{e}_n|$
are plotted in (a) and (b), respectively.
The different transitions $n$ are denoted by the different lines: $n=2$ (black solid), 3 (blue dashed), 4 (red dotted) and 5 (green dot-dash), respectively. 
Parameters $\gamma^{1n}_{eo}$ and $\gamma^{n1}_{eo}$ are plotted versus $U$ in (c)
and (d). The different transitions $n$ are denoted by the different symbols: $n=2$ (black circle), 3 (blue square), 4 (red upper triangle) and 5 (green diamond), respectively.
Parameters are used: $N=12$, $\mu=w$ and $\Delta=5w$.
}
\end{figure}

In Figs.~\ref{fig_high_freq_dissipation}(a) and (b), we plot the energy differences, $|\Delta^{eo}_{1n}|$ and $|\Delta^{eo}_{n1}|$, between 
the ground states and the first four eigen-energies
in their opposite parities, respectively. The energy difference decreases when the system exhibits the Majorana fermions, i.e., $\Delta{E}=0$ for $\Delta=5w$ in Fig.~\ref{Energy_gap1}.
Therefore, the mean number $\bar{n}_f(\Delta^{eo}_{nm})$ increases. Also,
it should be noted that the degeneracy between the higher
excited states occurs as shown in Figs.~\ref{fig_high_freq_dissipation}(a) and (b). This master
equation can still be used to describe the dissipative dynamics in the wide range of parameters except those
degeneracy points.

\begin{figure}[ht]
\centering
\includegraphics[height=8.0cm]{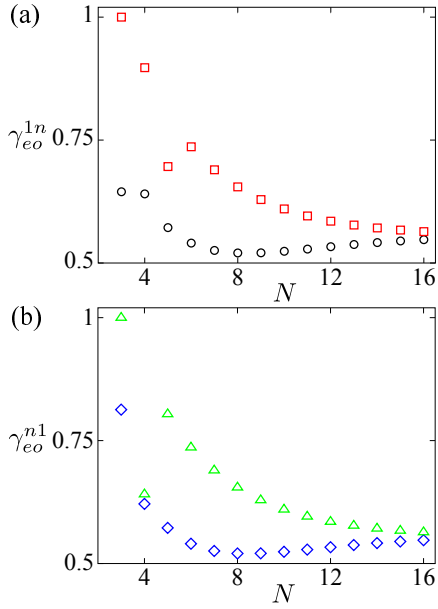}
\caption{ \label{high_freq_dissN} (Color online) 
Parameters $\gamma^{1n}_{eo}$ and $\gamma^{n1}_{eo}$ versus $N$, for $n=2,3$. In (a), $\gamma^{12}_{eo}$ and
$\gamma^{13}_{eo}$ are denoted by black circle and
red square, respectively. In (b), $\gamma^{21}_{eo}$ and
$\gamma^{31}_{eo}$ are denoted by blue diamond and
green upper triangle, respectively.
The parameters are used: $\mu=w$, $\Delta=5w$ and $U=8w$.
}
\end{figure}

The dissipation rate $\Gamma_{nm}$ is proportional
to $\sum_j|C^{nm}_j|^2$.  
Let us denote the parameters
$\gamma^{1n}_{eo}$ and $\gamma^{n1}_{eo}$ to be 
$\sum_j|C^{1n}_j|^2$ and
$\sum_j|C^{n1}_j|^2$, respectively. These parameters
give the transition rates between the ground state
and higher excited states in the opposite parity.
In Figs.~\ref{fig_high_freq_dissipation}(c) and (d), we plot the parameters $\gamma^{1n}_{eo}$ and $\gamma^{n1}_{eo}$ versus $U$, where $n=2,3,4$ and 5. These two parameters decreases when $U$ increases. Thus,
the dissipation rates $\Gamma_{1n}$ and $\Gamma_{n1}$
also decrease.
We can see that the interchange of the parameters $\gamma^{12}_\alpha$ and $\gamma^{13}_\alpha$ occurs around $U=7w$ in Figs.~\ref{fig_high_freq_dissipation}(c) and (d).
It is because the two energy levels avoid crossing 
around $U=7w$ in Figs.~\ref{fig_high_freq_dissipation}(a)
and (b), and the wavefunction must be continuous at this point.
Although the mean number $\bar{n}_f(\Delta^{eo}_{nm})$ increases as $U$ increases,
the parameters $\gamma^{1n}_{eo}$ and $\gamma^{n1}_{eo}$ decreases. Therefore, $\Gamma_{nm}\bar{n}_f$ decreases if the temperature $T$ is sufficiently low. The interaction
between fermions can reduce the effect of dissipation
at low temperature.

Also, we study the relationship between the behaviours
of $\gamma^{1n}_{eo}$ and $\gamma^{n1}_{eo}$ and the system's size. In Fig.~\ref{high_freq_dissN}, we plot the two parameters $\gamma^{1n}_{eo}$ and $\gamma^{n1}_{eo}$ versus $N$, for $n=2,3$. 
The parameters $\gamma^{12}_{eo}$ and $\gamma^{21}_{eo}$
decreases with small $N$, and then slightly increases
when $N$ becomes larger. The parameters $\gamma^{13}_{eo}$ and $\gamma^{31}_{eo}$ decrease with $N$. Besides, the parameters $\gamma^{12}_{eo}(\gamma^{12}_{oe})$ 
and $\gamma^{13}_{eo}(\gamma^{13}_{oe})$ start 
to converge at $N=16$ in Fig.~\ref{high_freq_dissN}.

\subsection{Dephasing}
We study the effect of dephasing on the Majorana modes.
In contrast to the case of dissipation, the dephasing noise gives rise to the transitions
between the eigenstates in the same parity. 
In this model, the fermions are coupled to a common
bosonic bath. The Hamiltonian, describes the system-bath coupling, is given by
\begin{eqnarray}
H_{BI}&=&\sum_{j}\tilde{g}_j{c^\dag_j{c_j}}B,
\end{eqnarray}
where $\tilde{g}_j$ is the coupling strength at site $j$ and $B$ is the bath operator. This decoherence model
is similar to the model discussed in \cite{Schmidt}.
Dephasing can induce the transitions between the eigenstates of the system which are summarized in Fig.~\ref{Decoherence_induced_transition}(b).

\begin{figure}[ht]
\centering
\includegraphics[height=5.0cm]{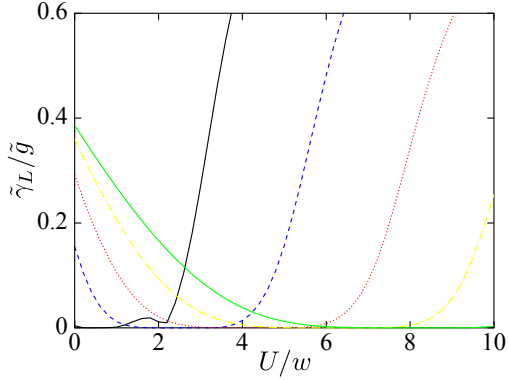}
\caption{ \label{low_freq_dephasing} (Color online) 
Parameter $\tilde{\gamma}_L$ versus interaction strength $U$, for $N=12$ and $\mu=w$. The different interaction strengths
$\Delta$ are denoted by the different lines: $\Delta=w$ (black solid), $2w$ (blue dashed), $3w$ (red dotted), $4w$ (yellow dash-dotted) and $5w$ (green solid), respectively. 
}
\end{figure}

\subsubsection{Low-frequency noise}
We study the effect of dephasing in the low-frequency
regime. In this regime, we can express this Hamiltonian in terms of eigenstates of
the two lowest degenerate states. Now the Hamiltonian is given by
\begin{eqnarray}
H_{BI}&=&\sum_j\tilde{g}_j(D^{11}_{ej}|1\rangle_e{}_e{\langle}1|+D^{11}_{oj}|2\rangle_o{}_o\langle{2}|)B(t),\\
&=&\frac{1}{2}\sum_j\tilde{g}_j(D^{11}_{ej}-D^{11}_{oj})(|1\rangle_e{}_e{\langle}1|-|2\rangle_o{}_o\langle{2}|)B(t)\nonumber\\
&&+\frac{1}{2}(D^{11}_{ej}+D^{11}_{oj})B(t),
\end{eqnarray}
where $\tilde{g}_j$ is the coupling strength, $D^{11}_{ej}$ and $D^{11}_{oj}$ are ${}_e\langle{1}|c^\dag_jc_j|{1}\rangle_e$
and ${}_o\langle{1}|c^\dag_jc_j|{1}\rangle_o$, respectively.
The effective coupling strength between the Majorana modes and bath is
\begin{eqnarray}
\tilde{\gamma}_L&=&\tilde{g}\Big|\sum_j(D^{11}_{ej}-D^{11}_{oj})\Big|,
\end{eqnarray}
where $\tilde{g}_l$ is roughly equal to $\tilde{g}$.
We study the relationship between the coupling strength $\tilde{\gamma}_L$ and the interaction strength $U$.
In Fig.~\ref{low_freq_dephasing}, we plot the parameter
$\tilde{\gamma}_L$ versus $U$, for the different strengths $\Delta$. The numerical results show
that $\tilde{\gamma}_L$ can reach nearly zero when the Majorana modes exhibit ($\Delta{E}=0$ in Fig.~\ref{Energy_gap1}). This shows that Majorana
modes are robust against the low-frequency dephasing noise. In fact, this can be easily understood by writing
the fermion operator in terms of spin operators. From Eq.~(\ref{occupation_number}), we have 
$c^\dag_jc_j=(\sigma^z_j+1)/2$. It will flip the spin
state from $|0\rangle_x(|1\rangle_x)$ to $|1\rangle_x(|0\rangle_x)$. It gives ${}_e\langle{1}|c^\dag_jc_j|{1}\rangle_e$ and ${}_o\langle{1}|c^\dag_jc_j|{1}\rangle_o$
to be 0.5 if the two degenerate ground states can be approximately described by the two N\'{e}el states.
Therefore, the parameter $\tilde{\gamma}_L$ is nearly zero.

\begin{figure}[ht]
\centering
\includegraphics[height=6.5cm]{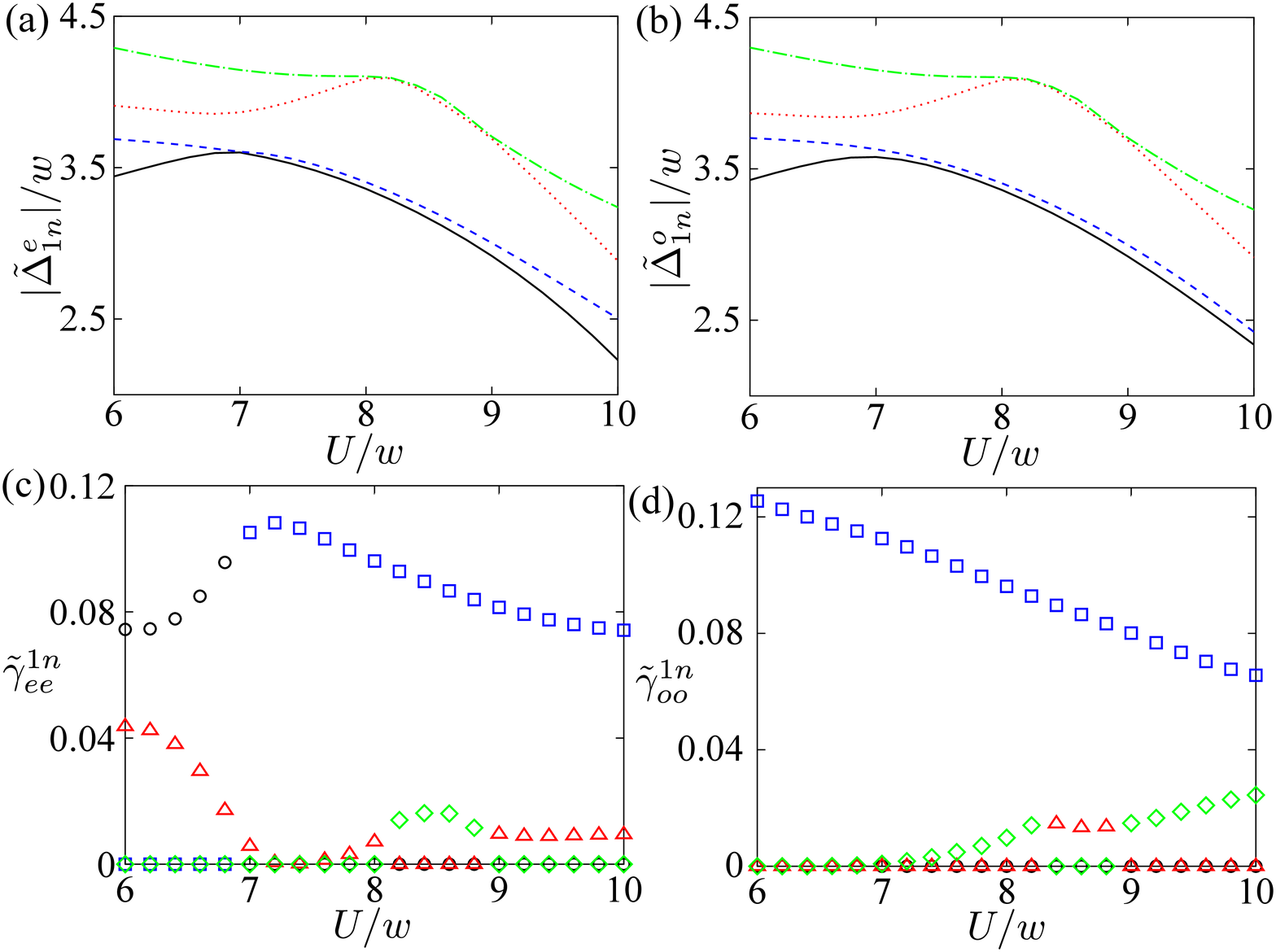}
\caption{ \label{fig_high_freq_dephasing} (Color online) 
Energy differences versus $U$ in (a)
and (b). The energy differences $|\tilde{\Delta}^{e}_{1n}|=|E^e_{1}-E^{e}_n|$ and $|\tilde{\Delta}^{o}_{1n}|=|E^o_{1}-E^{o}_n|$ are plotted in (a) and (b), respectively. The different transitions $n$ are denoted by the different lines: $n=2$ (black solid), 3 (blue dashed), 4 (red dotted) and 5 (green dot-dash), respectively.
In (c) and (d), the parameters $\tilde{\gamma}^{1n}_{ee}$ and $\tilde{\gamma}^{1n}_{oo}$ are plotted versus $U$. 
The different transitions $n$ are denoted by the different lines: $n=2$ (black circle), 3 (blue square), 4 (red upper triangle) and 5 (green diamond), respectively.
Parameters are used: $N=12$, $\mu=w$ and $\Delta=5w$.
}
\end{figure}

\subsubsection{High-frequency noise}
We consider the frequency domain of the noise
spectrum to be 
comparable to the transition frequency
between the different eigenstates. We presume 
that the high-frequency noise will not affect the dynamics between
the two degenerate ground states.
We follow the similar treatment in the previous
subsection to study the high-frequency noise.
We assume that the coupling between the system 
and bosonic bath is sufficiently weak, so
that the RWA can be applied.
In the interaction picture,
the Hamiltonian of system-bath coupling 
can be approximated as
\begin{equation}
H_{BI}=\sum_{j,k}\sum_{\alpha,n,m}\tilde{g}_{j}
D^{nm}_{\alpha{j}}|n\rangle_\alpha{}_\alpha\langle{m}|b_{j}e^{i(\tilde{\Delta}^\alpha_{nm}-\omega_{j})t}
+{H.c.},
\end{equation}
where $D^{nm}_{\alpha{j}}={}_\alpha\langle{n}|c^\dag_j{c_j}|{m}\rangle_\alpha$ and $\tilde{\Delta}^\alpha_{nm}=E^\alpha_n-E^\alpha_m$,
$\alpha=e,o$.
Here the energy difference $\tilde{\Delta}^\alpha_{nm}$
is positive and $n>m$.

The master equation can be obtained by using
the Born-Markovian approximation \cite{Beaudoin}. The master equation, describes the dephasing noise, can be written as
\begin{eqnarray}
\dot{\rho}&=&-i[H,\rho]+\sum_{nm}\tilde{\Gamma}^\alpha_{nm}\bar{n}_b(\tilde{\Delta}^\alpha_{nm})\mathcal{L}(|m\rangle_\alpha{}_\alpha\langle{n}|)\rho\nonumber\\
&&+\sum_{nm}\tilde{\Gamma}^\alpha_{nm}[1+\bar{n}_b(\tilde{\Delta}^\alpha_{nm})]\mathcal{L}(|n\rangle_\alpha{}_\alpha\langle{m}|)\rho,
\end{eqnarray}
where $\tilde{\Gamma}^\alpha_{nm}=2\pi\tilde{d}(\tilde{\Delta}^\alpha_{nm}){\tilde{g}^2}|\sum_j{D}^{nm}_{\alpha{j}}|^2$, 
$\tilde{\Omega}(\tilde{\Delta}^\alpha_{nm})$ is the density of states, $\tilde{g}_j\approx\tilde{g}$ and $n>m$. The parameter $\bar{n}_b(\tilde{\Delta}^\alpha_{nm})$ is the mean
occupation number, for the bosons, at the frequency $\tilde{\Delta}^\alpha_{nm}$
and the temperature $T$. Here we have assumed
that there is no degeneracy in the transitions \cite{Beaudoin}.

\begin{figure}[ht]
\centering
\includegraphics[height=6.0cm]{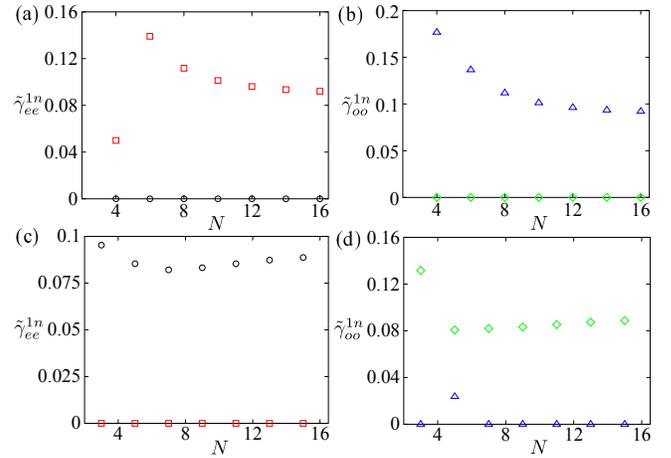}
\caption{ \label{High_freq_dephaseN} (Color online) 
Parameters $\tilde{\gamma}^{1n}_{ee}$ and $\tilde{\gamma}^{n1}_{oo}$ versus $N$, for $n=2,3$. 
The even number of fermions are plotted in
(a) and (b), and the odd number of fermions are plotted
in (c) and (d). 
In (a) and (c), $\tilde{\gamma}^{12}_{ee}$ and
$\tilde{\gamma}^{13}_{ee}$ are denoted by black circle and red square, respectively. In (b) and (d), $\tilde{\gamma}^{12}_{oo}$ and $\tilde{\gamma}^{13}_{oo}$ are denoted by green diamond and
blue upper triangle, respectively.
The parameters are used: $\mu=w$, $\Delta=5w$ and $U=8w$.
}
\end{figure}

In Figs.~\ref{fig_high_freq_dephasing}(a) and (b), 
we plot the energy differences $|\tilde{\Delta}^e_{1n}|$
and $|\tilde{\Delta}^o_{1n}|$ between
the ground state and the first four 
excited states in the same parity.
The energy difference decreases
when $U$ increases. The mean number
$\bar{n}_b(\tilde{\Delta}^\alpha_{nm})$
also increases with $U$. 
Then, we study the parameters
$\tilde{\gamma}^{1n}_{\alpha\alpha}=|\sum_j{D}^{1n}_{\alpha{j}}|^2$ as a function
of $U$. They are proportional to the dephasing
rate $\tilde{\Gamma}^{\alpha}_{nm}$.
In Figs.~\ref{fig_high_freq_dephasing}(c) and (d), we plot 
$\tilde{\gamma}^{1n}_{ee}$ and $\tilde{\gamma}^{1n}_{oo}$ versus $U$, where $n=2,3,4$ and 5. 
For even-parity transitions, the parameter $\tilde{\gamma}^{12}_{ee}$ increases and then
decreases when $U$ attain $7w$, and
$\tilde{\gamma}^{1n}_{ee}$ are much smaller than $\tilde{\gamma}^{12}_{ee}$ for higher $n$.
In the case of odd-parity transitions, the parameters
$\tilde{\gamma}^{12}_{oo}$ decreases when $U$ increases.
The parameter $\tilde{\gamma}^{13}_{oo}$ is nearly zero. 
However, $\tilde{\gamma}^{14}_{oo}$ increases as $U$ becomes larger. Since the energy difference between the ground state and the third and forth excited states are larger, this transition
is less important compared to the other transitions with the smaller energy gaps. The effect of dephasing, $\Gamma^\alpha_{nm}\bar{n}_b$, should be small if the temperature is sufficiently low. 

We also study the behaviours of the parameters 
$\tilde{\gamma}^{1n}_{ee}$ and $\tilde{\gamma}^{1n}_{oo}$, for the different system's sizes.
In Fig.~\ref{High_freq_dephaseN}, we plot the parameters $\tilde{\gamma}^{1n}_{ee}$
and $\tilde{\gamma}^{1n}_{oo}$ versus $N$. The results
are different for the even- and odd-number of fermions. The parameter $\tilde{\gamma}^{12}_{ee}$
is much smaller(larger) than $\tilde{\gamma}^{13}_{ee}$ in the even(odd)-number case. Similarly, $\tilde{\gamma}^{12}_{oo}$
is much smaller(larger) than $\tilde{\gamma}^{13}_{oo}$ if $N$ is even(odd).

\section{Discussion}
We have investigated the two general types of noises which are dissipation and dephasing, respectively. The low- and high-frequency noises are also discussed in each type of noise. Although we have not discussed the noise source for a specific environment, our study should capture the essential feature of the decoherence properties for various types of environment. We show the examples that long-range interactions between the fermions can change the decoherence properties of the Majorana modes. This is main result of our paper.

In addition, our study is related to the fundamental
problem in quantum mechanics. It is an important question
on the validity of quantum mechanics in the macroscopic
regime \cite{Schrodinger,Leggett}. Indeed, studies of macroscopic superpositions \cite{Ng2} shed light on this fundamental question \cite{Leggett}. One can consider to
create a superpositions of the two degenerate ground states of a fermonic chain which can be realized by either a 1D topological superconductor \cite{Kitaev} or trapped-ion chain \cite{Mezzacapo}. Although it is impossible to create the
superposition states of two Majorana fermions of a single
chain \cite{Goldstein,Rainis,Mazza} according to the superselection rule, it can be
resolved by encoding the states by using the four Majorana
fermions with two fermionic chains. We assume that decoherence does not set in between the two chains. Our present analysis can then be directly applied to this case. For a spin chain, the superposition of two degenerate
ground states can be created. The similar study can also be done. In fact, the fermionic and spin chains can be regarded as macroscopic systems. Thus, the decoherence properties of Majorana modes is important to understand the behavior of such superposition states.

\section{Conclusion}
In summary, we have studied the effect of dissipation and dephasing on the Majorana modes of a fermionic chain
in the presence of the nearest neighbor interactions between the fermions. The dissipation and dephasing 
noises can induce the parity- and non-parity preserving
transitions. We have also investigated the low- and
high-frequency noises to describe the different kinds
of environment. 
We show the examples that the dissipation and dephasing 
rates can be reduced by increasing the interaction strength at the sufficiently low temperature.
This means that the coherence time of Majorana fermions can be extended. It may be useful to the applications of QIP. In addition, we have studied the relationship between the decoherence rate and the system's size.

\begin{acknowledgments}
This work was supported in part by the 
National Basic Research Program of 
China Grants No. 2011CBA00300 and No.
2011CBA00301, the National Natural Science 
Foundation of China Grants No. 11304178, 
No. 61061130540, and No. 61361136003.
\end{acknowledgments}

\vspace{5mm}

\section*{Additional information}
Competing financial interests: The author declares no competing financial interests.

\appendix
\section{Spin-spin correlations}
\begin{figure}[ht]
\centering
\includegraphics[height=12cm]{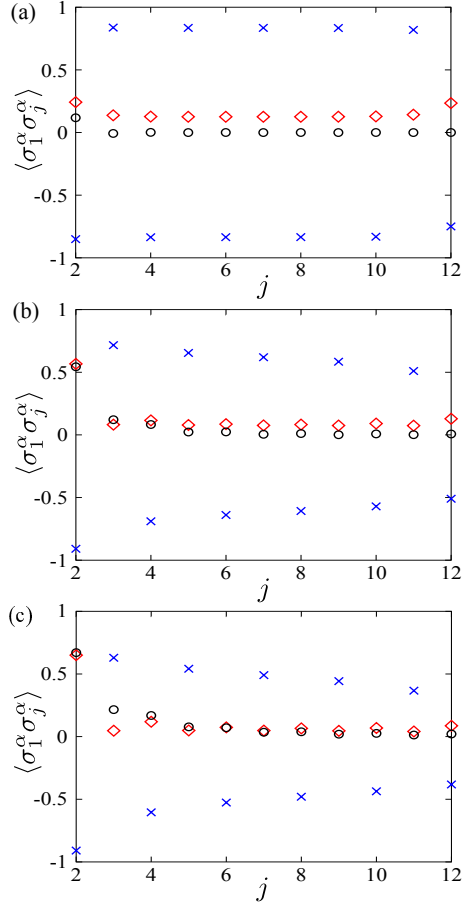}
\caption{ \label{spin-spin_corr} (Color online) Spin-spin correlations $\langle{\sigma^\alpha_1\sigma^\alpha_j}\rangle$
versus site $j$, where $N=12$,
$\mu=w$ and $U=0$. The different
strengths of $\Delta$ are plotted: (a) $\Delta=w$, 
(b) $\Delta=3w$ and (c) $\Delta=5w$. The different spin-spin correlations $\langle{\sigma^\alpha_1\sigma^\alpha_j}\rangle$ are denoted by $\alpha=x$ (blue cross),
$\alpha=y$ (black circle) and $\alpha=z$ (red diamond), respectively.
}
\end{figure}

The topological phase can be approximately described by the two N\'{e}el states in the $x$-direction, i.e.,
$|1010\ldots{10}\rangle_x$ and $|0101\ldots{01}\rangle_x$.
The properties of the topological phase can be manifested
by studying spin-spin correlations $\langle{{\sigma^\alpha_i}\sigma^\alpha_j\rangle}$, where $\alpha=x,y$ and $z$.
In Fig.~\ref{spin-spin_corr}, we plot the spin-spin correlations $\langle{{\sigma^\alpha_1}\sigma^\alpha_j\rangle}$
of the ground state versus site $j$, for the different strengths $\Delta$. When $\Delta=w$, spin-spin correlations $\langle{\sigma^x_1}\sigma^x_j\rangle$ alternate positive and negative ones with their
neighbouring spins in Fig.~\ref{spin-spin_corr}(a), and also they are about constant as the distance $|1-j|$ increases. The magnitude of correlations $\langle{\sigma^x_1}\sigma^x_j\rangle$ are larger than
the other correlations $\langle{\sigma^y_1}\sigma^y_j\rangle$
and $\langle{\sigma^z_1}\sigma^z_j\rangle$. This shows that the 
topological phase can be approximately described by the two N\'{e}el states in the $x$-direction.
As $\Delta$ increases, the interaction terms $\sigma^y_i\sigma^y_{i+1}$ appear. In Figs.~\ref{spin-spin_corr}(b)
and (c), the spin-spin correlations
$\langle{\sigma^x_1\sigma^x_j}\rangle$ are still dominant and show
alternating positive and negative numbers with their neighbours. However,
the correlations decays as the distance $|1-j|$ increases. This means that the N\'{e}el states are no longer a good approximation to describe the ground state.

\begin{figure}[ht]
\centering
\includegraphics[height=5.5cm]{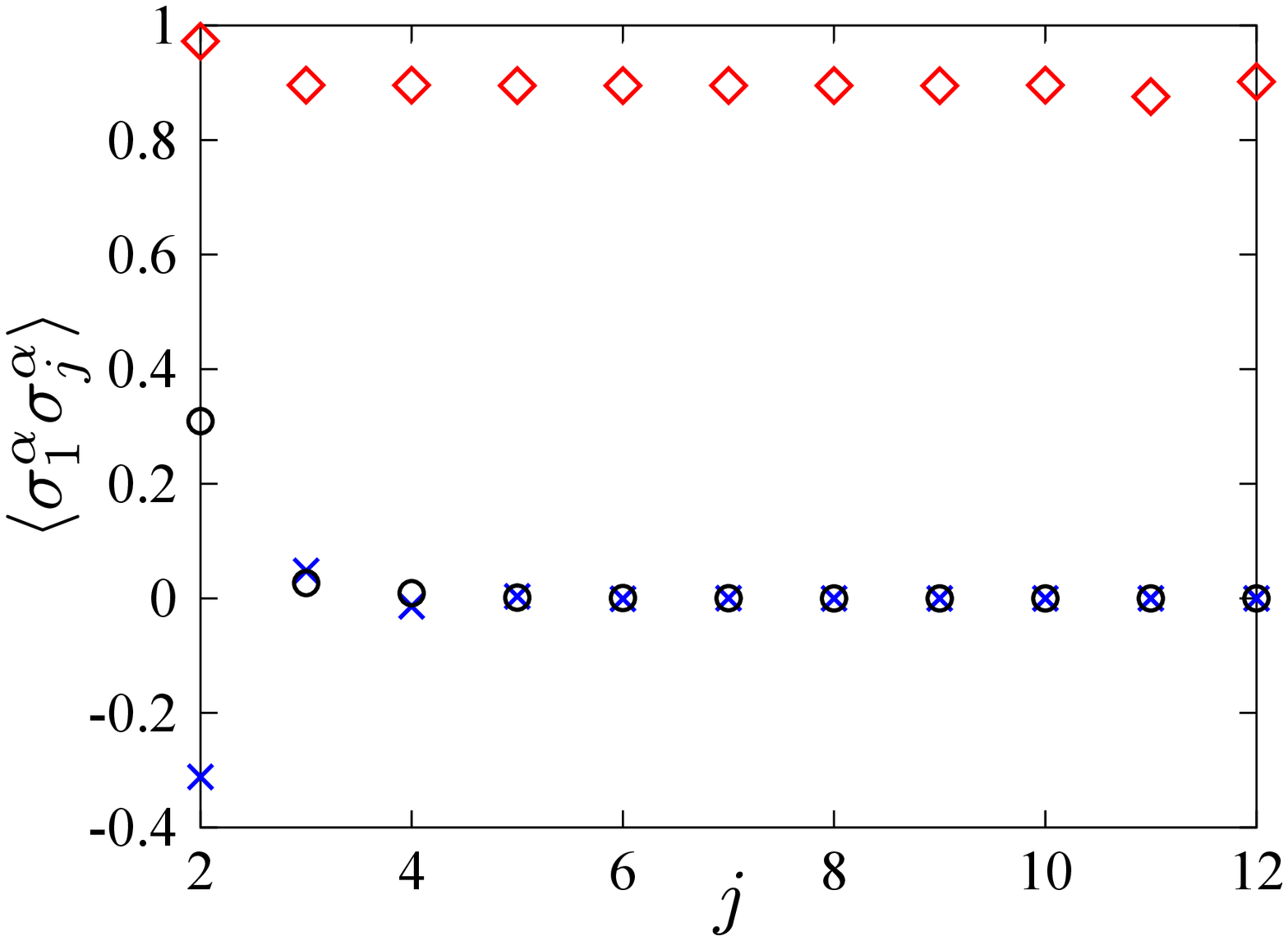}
\caption{ \label{PP} (Color online) Spin-spin correlations $\langle{\sigma^\alpha_1\sigma^\alpha_j}\rangle$
versus site $j$, where $N=12$, $\mu=10w$, $\Delta=5w$ and $U=-20w$. The different spin-spin correlations $\langle{\sigma^\alpha_1\sigma^\alpha_j}\rangle$ are denoted by $\alpha=x$ (blue cross),
$\alpha=y$ (black circle) and $\alpha=z$ (red diamond), respectively.
}
\end{figure}

We also study the spin-spin correlations of the ground states in the PP and AFM phases, respectively. In Fig.~\ref{PP}, we plot spin-spin correlations $\langle{{\sigma^\alpha_1}\sigma^\alpha_j\rangle}$ versus spin $j$, where the system is in the PP phase. In this case, the spins are polarized. Now $\langle{{\sigma^z_1}\sigma^z_j\rangle}$ are nearly one as the distance
between the spins increases, and $\langle{{\sigma^z_1}\sigma^z_j\rangle}$ are much larger than the other correlations
$\langle{{\sigma^x_1}\sigma^x_j\rangle}$ and $\langle{{\sigma^y_1}\sigma^y_j\rangle}$.

\begin{figure}[ht]
\centering
\includegraphics[height=5.5cm]{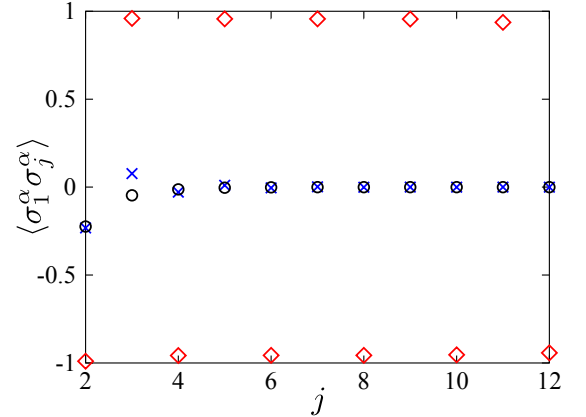}
\caption{ \label{AFM} (Color online) Spin-spin correlations $\langle{\sigma^\alpha_1\sigma^\alpha_j}\rangle$
versus site $j$, where $N=12$, $\mu=w$, $\Delta=5w$ and $U=20w$. The different spin-spin correlations $\langle{\sigma^\alpha_1\sigma^\alpha_j}\rangle$ are denoted by $\alpha=x$ (blue cross),
$\alpha=y$ (black circle) and $\alpha=z$ (red diamond), respectively.
}
\end{figure}

In Fig.~\ref{AFM}, we plot $\langle{{\sigma^\alpha_1}\sigma^\alpha_j\rangle}$ versus spin $j$, where the system is
in the AFM phase. We can see that the correlations in the $z$-direction are in alternating positive and negative ones with their neighbours. The other components of correlations are very small. This clearly shows that the ground state is in the AFM phase.

\section{Decoherence rate of low-frequency dissipative noise}
For the low-frequency dissipative noise,
the decoherence rate is closely related to the parameter
$\gamma_L$ which is given by
\begin{equation}
\gamma_L=g\Big|\sum_jC^{11}_j\Big|,
\end{equation}
and 
\begin{equation}
C^{11}_j={}_e\langle{1}|c_j+c^\dag_j|1\rangle_o.
\end{equation}
The numerical result shows that the parameter $\gamma_L$
does not show the dependence on the system size $N$.
This can be explained by considering the ground state in 
the topological phase. The two degenerate ground states can be approximately described by the superposition of two N\'{e}el states in the $x$-direction,
i.e., $|1010\ldots{10}\rangle_x$ and
$|0101\ldots{01}\rangle_x$.
The operator $c_j+c^\dag_j$ can be expressed in terms of the spin-half operators as
\begin{eqnarray}
c_j+c^\dag_j&=&(-1)^{j-1}\prod^{j-1}_{k=1}\sigma^z_k(\sigma^-_j+\sigma^+_j).
\end{eqnarray}
For example, we apply $c_1+c^\dag_1$ and $c_2+c^\dag_2$
to the state $|1010\ldots{10}\rangle_x$. We have
\begin{eqnarray}
(c_1+c^\dag_1)|1010\ldots{10}\rangle_x&=&(\sigma^-_1+\sigma^+_1)|1010\ldots{10}\rangle_x,\\
&=&|1010\ldots{10}\rangle_x,
\end{eqnarray}
and
\begin{eqnarray}
&&(c_2+c^\dag_2)|1010\ldots{10}\rangle_x,\\
&=&-\sigma^z_1(\sigma^-_2+\sigma^+_2)|1010\ldots{10}\rangle_x,\\
&=&(-1)(-1)|0010\ldots{10}\rangle_x,\\
&=&|0010\ldots{10}\rangle_x
\end{eqnarray}
The operator $\sigma^z_k$ will change
the state $|0\rangle_x(|1\rangle_x)$ to
$|1\rangle_x(|0\rangle_x)$. Similarly, the
product of operators $\prod^{j-1}_{k=1}\sigma^z_k$
will change the polarization of spin from $1$ to $j-1$.
Therefore, we have
\begin{eqnarray}
{}_x\langle{1010\ldots{10}}|(c_1+c^\dag_1)|1010\ldots{10}\rangle_x&=&1,\\
{}_x\langle{0101\ldots{01}}|(c_1+c^\dag_1)|0101\ldots{01}\rangle_x&=&-1
\end{eqnarray}
and 
\begin{eqnarray}
{}_x\langle{1010\ldots{10}}|(c_j+c^\dag_j)|1010\ldots{10}\rangle_x&=&0,\\
{}_x\langle{0101\ldots{01}}|(c_j+c^\dag_j)|0101\ldots{01}\rangle_x&=&0,\\
{}_x\langle{0101\ldots{01}}|(c_j+c^\dag_j)|1010\ldots{10}\rangle_x&=&0,
\end{eqnarray}
for $j\neq{1}$.
The parameter $C^{11}_1$ is about 1 and $C^{11}_{j\neq{1}}=0$ if
the ground state can be described by the superposition of two N\'{e}el states with equal weights.
Thus, the parameter $\gamma$ is close to $g$.


\begin{thebibliography}{99}
\bibitem{Wilczek}
Wilczek F. Majorana returns. \textit{Nat. Phys.} {\bf 5}, 614 (2009).



\bibitem{Read}
Read N. \& Green D. Paired states of fermions in two dimensions with breaking of parity and time-reversal symmetries and the fractional quantum Hall effect. \textit{Phys. Rev. B} {\bf 61}, 10267 (2000).

\bibitem{Ivanov}
Ivanov D. A. Non-Abelian Statistics of Half-Quantum Vortices in $p$-Wave Superconductors. \textit{Phys. Rev. Lett.} {\bf 86}, 268 (2001).

\bibitem{Alicea}
Alicea J., Oreg Y., Refael G., von Oppen F. \&
Fisher M. P. A. Non-Abelian statistics and topological quantum information processing in 1D wire networks. \textit{Nat. Phys.} {\bf 7}, 412 (2011).

\bibitem{Nayak}
Nayak C., Simon S. H., Stern A., Freedman M. \& Das Sarma S. Non-Abelian anyons and topological quantum computation. \textit{Rev. Mod. Phys.} {\bf 80}, 1083 (2008).

\bibitem{Kitaev}
Kitaev A. Y. Unpaired Majorana fermions in quantum wires. \textit{Physics-Uspekhi} {\bf 44}, 131 (2001).

\bibitem{Lutchyn}
Lutchyn R. M., Sau J. D. \& Das Sarma S. Majorana Fermions and a Topological Phase Transition in Semiconductor-Superconductor Heterostructures. 
\textit{Phys. Rev. Lett.} {\bf 105}, 077001 (2010).

\bibitem{Oreg}
Oreg Y., Refael G. \& von Oppen F. Helical Liquids and Majorana Bound States in Quantum Wires. \textit{Phys. Rev. Lett.} {\bf 105}, 177002 (2010). 

\bibitem{Kraus}
Kraus C. V., Diehl S., Zoller P. \& Baranov M. A.
Probing Atomic Majorana Fermions in Optical Lattices.
\textit{New J. Phys.} {\bf 14}, 113036 (2012).

\bibitem{Jiang}
Jiang L. {\it et al.} Majorana Fermions in Equilibrium and in Driven Cold-Atom Quantum Wires. \textit{Phys. Rev. Lett.} {\bf 106}, 220402 (2011).


\bibitem{Nielsen}
Nielsen M. \& Chuang I. L. {\it Quantum Computation
and Quantum Information}, (Cambridge University Press, Cambridge, 2001).

\bibitem{Shor}
Shor P. Scheme for reducing decoherence in quantum computer memory. \textit{Phys. Rev. A} {\bf 52}, R2493 (1995).

\bibitem{Steane}
Steane A. M. Error Correcting Codes in Quantum Theory. \textit{Phys. Rev. Lett.} {\bf 77}, 793 (1996).

\bibitem{Viola}
Viola L. \& Lloyd S. Dynamical suppression of decoherence in two-state quantum systems. \textit{Phys. Rev. A} {\bf 58}, 2733 (1998).

\bibitem{Facchi}
P. Facchi P. {\it et al.} Control of decoherence: Analysis and comparison of three different strategies. \textit{Phys. Rev. A} {\bf 71}, 022302 (2005).

\bibitem{Kitaev2}
Kitaev A. \& Laumann C. Topological phases and quantum computation. \textit{arXiv preprint} (2009) 0904.2771.



\bibitem{Goldstein}
Goldstein G. \& Chamon C. Decay rates for topological memories encoded with Majorana fermions. \textit{Phys. Rev. B} {\bf 84}, 205109 (2011).

\bibitem{Budich}
Budich J. C., Walter S. \& Trauzettel B. Failure of protection of Majorana based qubits against decoherence. \textit{Phys. Rev. B} {\bf 85}, 121405(R) (2012).

\bibitem{Schmidt}
Schmidt M. J., Rainis D. \& Loss D. Decoherence of Majorana qubits by noisy gates. \textit{Phys. Rev. B} {\bf 86}, 085414 (2012).

\bibitem{Cheng}
Cheng M., Lutchyn R. M. \& Das Sarma S. Topological protection of Majorana qubits. \textit{Phys. Rev. B} 85, 165124 (2012).

\bibitem{Rainis}
Rainis D. \& Loss D. Majorana qubit decoherence by quasiparticle poisoning. \textit{Phys. Rev. B} {\bf 85}, 174533 (2012).




\bibitem{Ho}
Ho S.-H., Chao S.-P., Chou C.-H. \& Lin F.-L. Decoherence Patterns of Topological Qubits from Majorana Modes. \textit{New J. Phys.} \textbf{16} 113062 (2014).


\bibitem{Stoudenmire}
Stoudenmire E. M., Alicea J., Starykh O. A. \& Fisher M. P. A. Interaction effects in topological superconducting wires supporting Majorana fermions. \textit{Phys. Rev. B} {\bf 84}, 014503 (2011).

\bibitem{Gangadharaiah}
Gangadharaiah S., Braunecker B., Simon P. \& Loss D. Majorana Edge States in Interacting One-Dimensional Systems. \textit{Phys. Rev. Lett.} {\bf 107}, 036801 (2011).

\bibitem{Sela}
Sela E., Altland A. \& Rosch A. Majorana fermions in strongly interacting helical liquids. \textit{Phys. Rev. B} {\bf 84}, 085114 (2011).

\bibitem{Hassler}
Hassler F. \& Schuricht D. Strongly interacting Majorana modes in an array of Josephson junctions. \textit{New J. Phys.} {\bf 14}, 125018 (2012).

\bibitem{Thomale}
Thomale R., Rachel S. \& Schmitteckert P. Tunneling spectra simulation of interacting Majorana wires. \textit{Phys. Rev. B} \textbf{88}, 161103(R) (2013).

\bibitem{Ng}
Ng H. T. Topological phases in spin-orbit-coupled dipolar lattice bosons. \textit{Phys. Rev. A} \textbf{90}, 053625 (2014).

\bibitem{Ladd}
Ladd T. D. {\it et al.} Quantum computers. \textit{Nature} {\bf 464}, 45 (2010).

\bibitem{Paladino}
Paladino E., Galperin Y. M., Falci G. \& Altshuler B. L. 1/f noise: Implications for solid-state quantum information. \textit{Rev. Mod. Phys.} {\bf 86}, 361 (2014).



\bibitem{Mazza}
Mazza L., Rizzi M., Lukin M. D. \& Cirac J. I. Robustness of quantum memories based on Majorana zero modes. \textit{Phys. Rev. B} {\bf 88}, 205142 (2013).

\bibitem{Beaudoin}
Beaudoin F., Gambetta J. M. \& Blais A. Dissipation and ultrastrong coupling in circuit QED. \textit{Phys. Rev. A} {\bf 84}, 043832 (2011).

\bibitem{Mezzacapo}
Mezzacapo A., Casanova J., Lamata L. \& Solano E.
Topological qubits with Majorana fermions in trapped ions. \textit{New J. Phys.} \textbf{15}, 033005 (2013).


\bibitem{Pinheiro}
Pinheiro F., Bruun G. M., Martikainen J.-P. \& Larson J. $XYZ$ Quantum Heisenberg Models with $p$-Orbital Bosons. \textit{Phys. Rev. Lett.} \textbf{111}, 205302 (2013).




\bibitem{Breuer}
Breuer H. P. \& Petruccione F. {\it The Theory of 
Open quantum systems}, (Oxford University Press, New York, 2007).

\bibitem{Schrodinger}
Schr\"{o}dinger E. Die gegenw\"{a}rtige Situation in der Quantenmechanik. \textit{Die Naturwissenschaften} \textbf{23}, 807 (1935).

\bibitem{Leggett}
Leggett A. J. Testing the limits of quantum mechanics: motivation, state of play, prospects. \textit{J. Phys.: Condens. Matter} \textbf{14}, R415 (2002).

\bibitem{Ng2}
Ng H. T. Production of mesoscopic superpositions with ultracold atoms. \textit{Phys. Rev. A} {\bf 77}, 033617 (2008).

\end{thebibliography}
\end{document}